# An evolutionary medicine and life history perspective on aging and disease: Trade-offs, hyperfunction, and mismatch

Jacob E. Aronoff & Benjamin C. Trumble

School of Human Evolution and Social Change, Center for Evolution and Medicine, Institute of Human Origins, Arizona State University, Tempe, Arizona, USA

**Heading Title:** A life history perspective on chronic disease

**Author contact information:**
Jacob E. Aronoff (correspondence)
Center for Evolution & Medicine
LSC 224 Tempe, AZ 85287-4501, USA
Phone: 616-901-3143
Email: jarnoff@asu.edu

Benjamin C. Trumble
Center for Evolution & Medicine
LSC 218 Tempe, AZ 85287-4501, USA
Phone: 480-965-1394
Email: trumble@asu.edu

**Abstract**

The rise in chronic diseases over the last century presents a significant health and economic burden globally. Here we apply evolutionary medicine and life history theory to better understand their development. We highlight an imbalanced metabolic axis of growth and proliferation (anabolic) versus maintenance and dormancy (catabolic), focusing on major mechanisms including IGF-1, mTOR, AMPK, and Klotho. We also relate this axis to the hyperfunction theory of aging, which similarly implicates anabolic mechanisms like mTOR in aging and disease. Next, we highlight the Brain-Body Energy Conservation model, which connects the hyperfunction theory with energetic trade-offs that induce hypofunction and catabolic health risks like impaired immunity. Finally, we discuss how modern environmental mismatches exacerbate this process. Following our review, we discuss future research directions to better understand health risk. This includes studying IGF-1, mTOR, AMPK, and Klotho and how they relate to health and aging in human subsistence populations, including with lifestyle shifts. It also includes understanding their role in the developmental origins of health and disease as well as the social determinants of health disparities. Further, we discuss the need for future studies on exceptionally long-lived species to understand potentially underappreciated trade-offs and costs that come with their longevity. We close with considering possible implications for therapeutics, including (1) compensatory pathways counteracting treatments, (2) a "Goldilocks zone", in which suppressing anabolic metabolism too far introduces catabolic health risks, and (3) species constraints, in which therapeutics tested in shorter lived species with greater anabolic imbalance will be less effective in humans.

**Life History Theory, a metabolic axis, and disease**

With lifestyle shifts and increasing life expectancy, the last century has seen a dramatic rise in prevalence of age-related non-communicable diseases (NCD). These diseases now account for the majority of deaths globally at 43 million (~75%) (1, 2). In addition to their health burden, NCD have come with significant financial cost. For high income countries, this has put significant strain on the health care system (3), while among low- and middle-income countries (LMIC) it is a major contributor to poverty (4). Understanding, preventing, and treating these diseases is therefore one of the most pressing health and economic problems of modernity.

Here we utilize evolutionary medicine and life history theory (LHT) as an overarching framework for understanding NCD. LHT is a framework in evolutionary biology aimed at explaining how natural selection shapes an organism's lifecycle to maximize survival and reproduction, including growth and lifespan. A major focus of LHT research is studying how organisms utilize and allocate energy (5). For example, under the favorable environmental condition of caloric abundance, an organism is expected to invest in anabolic processes of growth and proliferation, while short-term caloric scarcity is expected to activate catabolic processes of maintenance and dormancy (6). Optimally regulating this axis in response to fluctuating environmental conditions is critical. However, recent developments in the evolutionary theories of aging, in particular the hyperfunction theory, suggest its optimization for early life growth and development comes at the expense of optimizing for longevity (7-9). Further, this later life imbalance is accelerated and exacerbated by modern environments with unprecedented caloric excess and sedentary lifestyles (6, 10). Several chronic diseases that increase with age and are commonly found in modern industrialized populations (e.g., cardiovascular diseases (CVD), type 2 diabetes, certain cancers, autoimmunity, chronic kidney disease (CKD), non-alcoholic fatty liver disease (NAFLD), Alzheimer's Disease and Related Dementias (ADRD)), share a common profile: over-activation of anabolic growth/proliferation pathways and under-activation of catabolic maintenance/dormancy (11-16).

**Major mechanisms of the anabolic-catabolic axis and their links to development and disease**

*Anabolic growth and proliferation: GH/IGF-1 and mTOR*

Insulin-like growth factor 1 (IGF-1) and mammalian/mechanistic target of rapamycin (mTOR) have received considerable attention for their roles in aging and disease. From an evolutionary LHT perspective, they are part of a nutrient sensing system that activates anabolic processes of growth/proliferation during periods of abundance, with IGF-1 being one important upstream regulator of mTOR (15, 17). In addition to being activated by growth factors like IGF-1, mTOR is also activated by other signals of energy and nutrient status like amino acids, glucose, and insulin. While mTOR is a serine/threonine protein kinase consisting of two complexes (mTORC1 and mTORC2), we restrict our discussion to complex 1, which plays a more central role in growth and proliferation (12). In addition, while many studies have focused on IGF-1, it is important to note that growth hormone (GH) plays a critical stimulatory role in the IGF-1 pathway (18) (**Figure 1**). As a result, we will refer to the GH/IGF-1 pathway unless referencing a study specifically measuring IGF-1.

The GH/IGF-1 and mTOR pathways serve critical functions in early life growth and development. During pregnancy, their activation in the placenta augments fetal growth in response to maternal nutrient availability (19-22). There is evidence that mTOR suppression is a mediating pathway linking childhood malnutrition to stunting (23, 24), while lower GH/IGF-1 due to genetic mutation has been found to result in shorter stature and delayed puberty (25). mTOR also plays an important role in brain development, including proliferation and differentiation of neurons and glia (26, 27).

The GH/IGF-1 and mTOR pathways also play critical roles in reproductive functioning, and their suppression can lead to infertility (28, 29). IGF-1 is found in seminal plasma and improves sperm motility (28). It also improves ovarian function and endometrial receptivity (30). Further, mTOR augments spermatogenesis, follicle development, oocyte meiotic maturation, and placental development and implantation (22, 29, 31).

In later life, high circulating levels of IGF-1 have been associated with increased risk for several chronic diseases, including CVD, cancer, type 2 diabetes, and ADRD as well as all-cause mortality (32). Genetic studies, including Mendelian randomization and rare genetic variants, have provided causal evidence linking elevated GH/IGF-1 signaling to cancer, type 2 diabetes, and coronary artery disease (33). Similarly, heightened mTOR activation has been implicated in several chronic diseases, such as type 2 diabetes, certain cancers, CVD, ADRD, CKD, and NAFLD (12, 13, 16, 17). mTOR-mediated cellular growth and inflammation in the kidney is one potential mechanism contributing to CKD (12), while mTOR also augments growth and proliferation of cancer cells (13). More broadly, mTOR stimulates mitochondrial biogenesis, and with chronic activation this can lead to excess reactive oxygen species (ROS) production contributing to cellular damage (13, 34). Another major mechanism through which the GH/IGF-1 and mTOR pathways can increase health risk is through inhibiting autophagy, as we will highlight below (13, 35).

mTOR has also been implicated in autoimmunity (36). While immune function is typically categorized as part of maintenance in life history research, this categorization is based on the response to infections, which is energetically costly and can come at the expense of early life growth or reproduction (37, 38). To clarify, we place this type of immune function as part of growth/proliferation, following Wang and colleagues (2019), because it requires mTOR-mediated immune cell proliferation (6). mTOR shifts the balance of helper T cells in favor of $T_H1/T_H17$, downregulating anti-inflammatory Tregs (39). This process could accelerate the loss of host tolerance, as $T_H17$ are involved in the pathogenesis of autoimmunity (40). In addition, the effect of mTOR activation on the helper T cell profile could reflect an immune response intended to increase surveillance of host cells, potentially preventing tumorigenesis in a pro-growth/proliferation environment. mTOR activation in T cells can improve cancer surveillance (17). Further, an inverse association between autoimmunity and cancer prevalence has been highlighted (41), while therapeutic treatments for cancer can often induce autoimmunity (42).

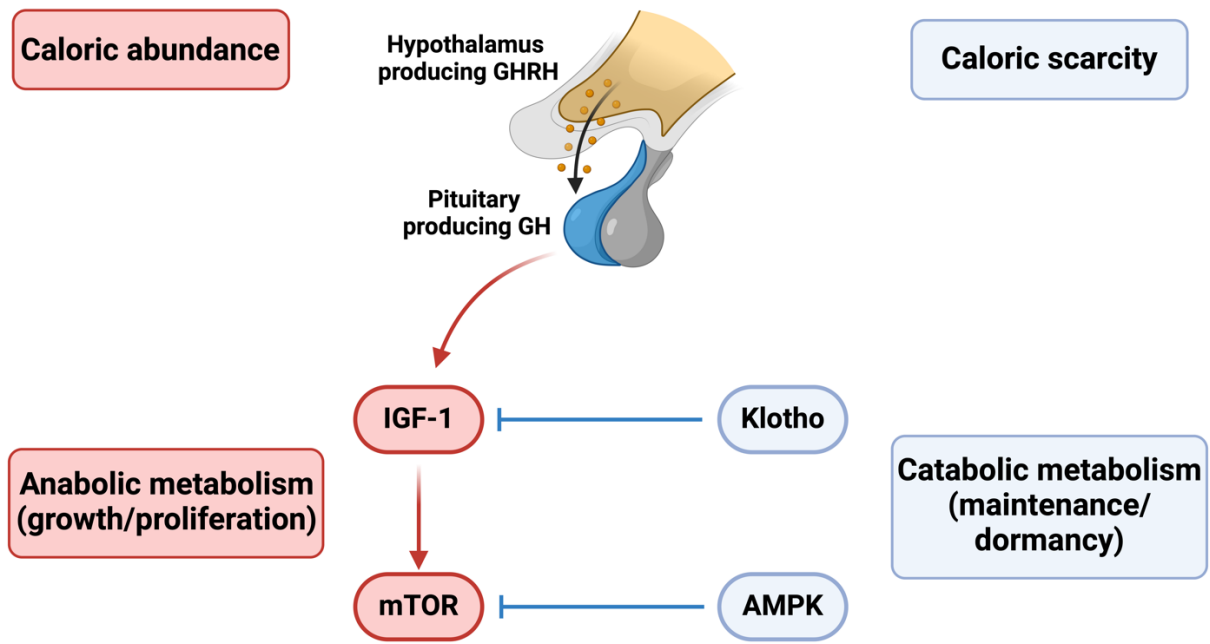

**Figure 1.** The relationship between GH/IGF-1, mTOR, Klotho, and AMPK, including how they map onto energy status and metabolism.

*Catabolic maintenance and dormancy: AMPK and Klotho*

AMP-activated protein kinase (AMPK) is a serine/threonine protein kinase and energy sensing enzyme that works antagonistically with IGF-1 and mTOR, and its action has been implicated in reduced disease risk (15). It is an activator of catabolic metabolism, including beta oxidation and glycolysis. In response to scarcity (low ATP), AMPK suppresses mTOR and activates autophagy, a process of cellular maintenance that involves degrading misfolded proteins, damaged organelles, and other abnormal cell components (43). AMPK also plays an important role in regulating growth/proliferation in response to fluctuating energy status in early life. For example, an experimental rat study found that greater AMPK expression in the hypothalamus following malnutrition played a role in delaying puberty onset (44). This is consistent with the use of the AMPK-activator metformin in humans. In addition to being used in the treatment of type 2 diabetes, CVD, certain cancers, and ADRD, metformin has also been used to prevent precocious puberty (45, 46). Experimental evidence in mice suggests the ability to inhibit mTOR and activate autophagy is likely critical for surviving through periods of energy scarcity, as neonates without this ability could not survive prolonged fasting (47).

Klotho is a gene and protein that has received considerable attention recently as an anti-aging target, although more specifically its functioning falls into catabolic maintenance/dormancy. Klotho inhibits IGF-1 and activates autophagy, while increased levels have been observed with mTOR inhibitors such as rapamycin and metformin (48, 49). Klotho levels are inversely associated with chronic diseases such as CKD, type 2 diabetes, hypertension, and ADRD, while preclinical Klotho therapy has been found to improve these conditions (48, 50). In the case of ADRD for example, the clearance of proteins such as amyloid-β through autophagy has been proposed to explain the neuroprotective effect of Klotho (51). There is also evidence this neuroprotective effect operates through platelets, as well as other possible pathways not yet known (52).

*What about anabolic maintenance?*

A complication with the growth/proliferation versus maintenance/dormancy terminology is that somatic maintenance involves both anabolic and catabolic processes. In the context of immune strategies, this has been categorized as defense (anabolic), involving mTOR-mediated immune cell proliferation to combat infection, versus dormancy (catabolic), or tolerance (6). More broadly, anabolic maintenance includes cellular repair and tissue remodeling in addition to immune defense. Since these are important processes for understanding aging, we specify anabolic versus catabolic maintenance in the sections below.

**Life history theory and the hyperfunction theory of aging**

The anabolic-catabolic life history axis aligns with programmatic theories of aging, most notably the hyperfunction theory proposed by Blagosklonny (7, 53, 54). While this theory is not new (8), it is gaining greater acceptance in recent years due to its ability to explain links between mTOR inhibition and longevity (7). According to the theory, the functioning of anabolic pathways like GH/IGF-1 and mTOR were optimized for early life growth and development. However, due to the selection shadow in later life, their activation is too high (hyperfunctional), which contributes to aging and disease (7-9, 53). Similar to the Disposable Soma Theory of aging (DST), the hyperfunction theory is an antagonistic pleiotropy model, in which traits are selected that confer early life fitness benefits but come with later life costs (55). However, the DST presents a trade-off between energy invested in reproduction at the cost of anabolic maintenance, which contributes to the accumulation of somatic damage (56). In contrast, the hyperfunction theory presents a functional trade-off between anabolic and catabolic cellular metabolism (**Figure 2**) (55).

Support for the hyperfunction theory has come from multiple lines of evidence. Most notably, it explains the relationship between mTOR suppression (reduced hyperfunction) and lifespan extension of laboratory mice and other species (7, 53). In addition, cross-species comparisons have indicated that natural selection on the GH/IGF-1 pathway helps explain variation in longevity (57, 58). Within humans, studies of centenarians have also reported evidence for genetic variation in GH/IGF-1 functioning (59, 60). Epigenetic clocks, which capture biological aging through variation in DNA methylation, have found consistent age-related changes across mammalian tissues in proximity to genes involved in development (61, 62). Finally, senescent cells, which accumulate with age and contribute to dysfunction and disease, show characteristics of hyperfunction (53). Under normal functioning, these cells are involved in anabolic maintenance through tissue remodeling (63). However, in later life they display greater mTOR expression as well as pro-inflammatory signaling, which has been termed the senescent-associated secretory profile (SASP) (53, 63-65). SASP cells are a major contributor to "inflammaging", the development of chronic low-grade inflammation that is both a cause and consequence of aging and disease (64).

The hyperfunction and DST models are not mutually exclusive, as both hyperfunction and damage accumulation are part of a multifactorial process (7, 9). Further, they are likely synergistic, as damage accumulation is a major contributor to SASP development (64). These cells in turn can impair and induce damage to surrounding cells, triggering a vicious cycle (66-68).

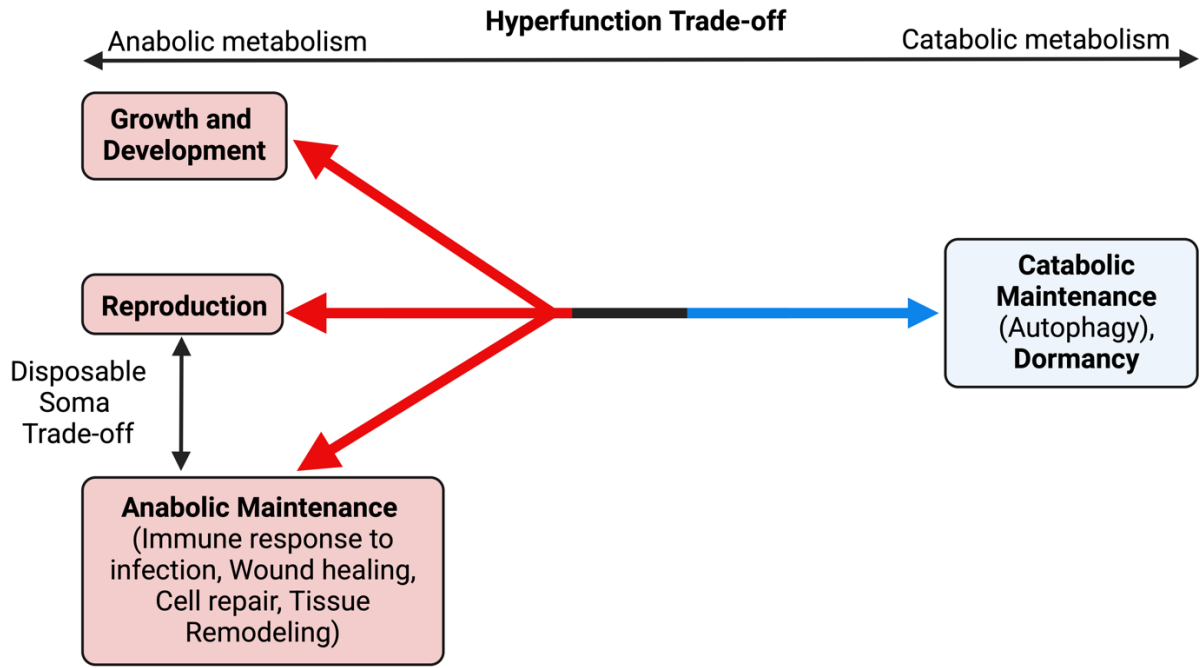

**Figure 2.** Distinguishing between anabolic energy trade-offs and the anabolic-catabolic functional trade-off, including how they relate to the DST and hyperfunction theories of aging.

*What about **hypo**function?*

A potential complication of the hyperfunction theory is that late life is also characterized by *hypo*function that contributes to aging and health risk (55). This is reflected for example in the loss of bone and muscle tissue (osteoporosis and sarcopenia respectively), brain atrophy, and impaired immune responses to infection (64). It is well-documented that the mTOR pathway plays an important role in the immune response, including the differentiation, activation, and function in T cells, B cells, and antigen-presenting cells (6, 69). Further, it is not only high IGF-1 but low levels as well that are associated with increased risk for most chronic diseases and mortality (14, 32). Hypofunction of the GH/IGF-1 and mTOR pathways can also increase later life health risk through impaired recovery from injuries (38, 70). Animal models indicate they play a role in central nervous system repair, which likely involves inhibiting autophagy (71-73).

They are also critical for muscle regeneration (74, 75). In an extreme example, the capacity to regenerate limbs following amputation in the axolotl (*Ambystoma mexicanum*) is mTOR-mediated (76). There is also evidence the GH/IGF-1 pathway contributes to tissue remodeling and improved cardiac functioning following myocardial infarction in humans (77).

As infection and injury have been the major causes of death throughout human history (78, 79), hypofunction in later life presents a seeming paradox. However, there are two mechanisms through which *hyper*function possibly contributes to *hypo*function. The first is the accumulation of SASP cells, which contribute to hypofunction by dysregulating, damaging, and overall impairing surrounding cells (66, 67). The second is through coordinated energy trade-offs, as outlined in the Brain-Body Energy Conservation (BEC) model (64). This model is based on a core tenet of LHT, that organisms have finite energy budgets from which to allocate across competing functions, resulting in trade-offs (80). Hyperfunctional SASP cells signal increased energy expenditure to the brain, in particular the hypothalamus, which orchestrates compensatory divestment from functions not immediately necessary for survival. This includes for example the immune repertoire, reflected in thymus involution and decreased naïve T cells, which only provides future benefits after encountering new pathogens (64). These processes therefore create a synergy of damage, hyperfunction, and hypofunction, which progressively leads to functional decline, disease, and death (**Figure 3**).

*Hyper*function-driven *hypo*function can explain the seemingly paradoxical effects of rapamycin on immune function. Despite its direct immunosuppressive effect by inhibiting mTOR, rapamycin administration has been found to boost antibody responses to vaccination and improve responses to infection over follow up (81, 82). This immune-boosting effect appears to occur indirectly, through rapamycin suppressing SASP cells (53, 83). As a result, their hypermetabolic signaling to the brain decreases, alleviating compensatory divestment from the immune repertoire and resulting in a stronger immune response to novel infectious exposures (64).

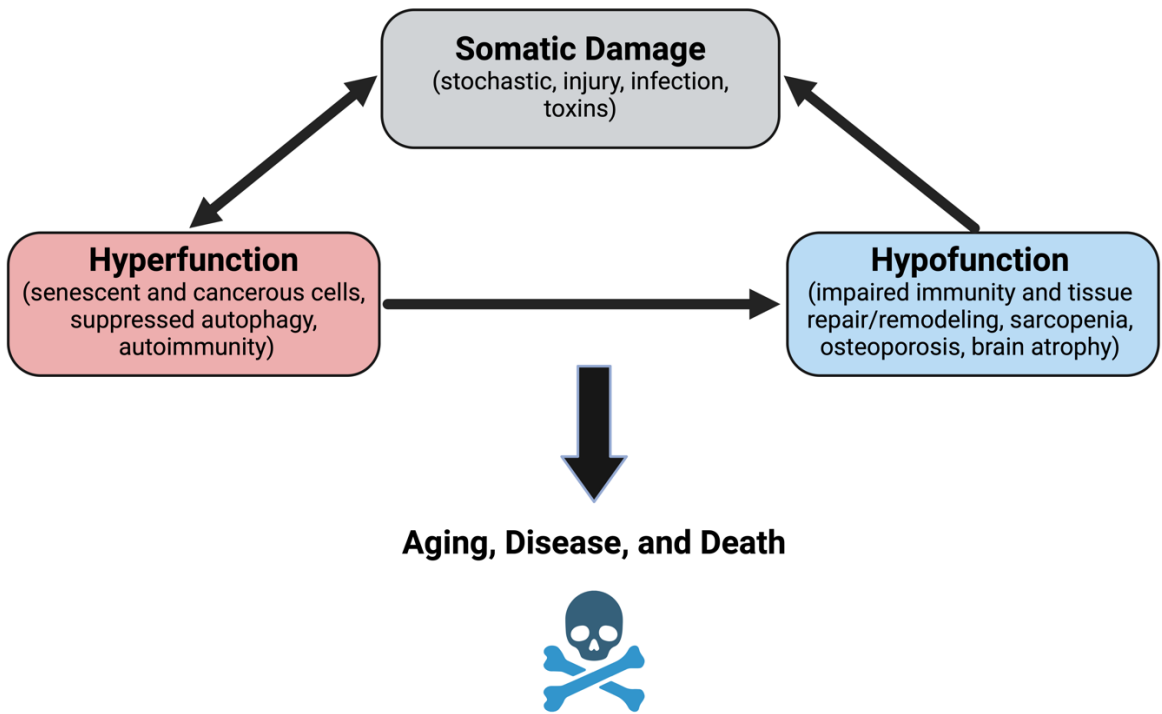

**Figure 3.** The synergies of hyperfunction, damage, and hypofunction that contribute to aging, disease, and death.

**Environmental mismatch: Speeding up the car and driving it off the road**

It is well-recognized that modern industrialized lifestyles of caloric surplus and limited physical activity present an environmental mismatch with our evolved biology that contributes to disease (10). Modern environments contribute to anabolic imbalance, leading to both earlier development of age-related NCD as well as new manifestations of dysregulation and disease. In Blagosklonny's description of his hyperfunction theory, he used the analogy of a car driving at an appropriate speed on the highway (early life), but too fast in the driveway (later life) (9). Extending this analogy, modern environmental mismatches speed up the car. This can be seen for certain diseases when comparing US or European samples with a subsistence population like the Tsimane forager-horticulturalists in the Bolivian Amazon. The Tsimane develop cardiovascular disease at a much slower rate, and as a result have among the lowest prevalence ever reported (84). Similarly, brain atrophy with age occurs at a significantly slower rate compared to US and European samples (85).

In other cases, modern environments do not simply speed up aging but create new manifestations of hyperfunctional anabolic imbalance. This is more analogous to driving the car off the road. For example, with industrialization has come increased exposure to air pollution and cigarette smoke. This contributes to chronic lung damage and inflammation, which can develop into chronic obstructive pulmonary disease (COPD) and cancer (79). Manifestations of anabolic imbalance can also be seen in early life. For example, hyperfunction of mTOR has been implicated in both epilepsy and autism, with rapamycin currently being used as a promising therapeutic (86-88).

**Future research directions**

*Why do individuals vary along the anabolic-catabolic spectrum? What are the trade-offs and health consequences?*

This evolutionary medicine and life history theory framework raises several questions that can be addressed by future studies (**Table 1**). From a cross-species perspective, humans are relatively long lived. We are also slower to develop and reach maturity compared to other primates, fitting the hyperfunction theory of aging. Comparative research across human subsistence populations and primate species suggests this is likely due to the long period of time needed for brain development to acquire complex foraging skills (89). In further support of this connection, human developmental changes in the cerebral metabolic rate of glucose track inversely with growth (90). However, there is also individual variation in aging due partly to genetic background (59, 60). This suggests a trade-off involving costs and benefits along the anabolic-catabolic spectrum. A similar dynamic has been observed with the APOE4 genotype. While carrying this allele increases risk for ADRD, studies in populations experiencing a high pathogen load have found it can be protective for cognitive development and functioning (91, 92). Further, Tsimane women with the APOE4 genotype have a larger number of offspring on average, due to a combination of reaching maturity earlier and having a shorter interbirth interval (91). Studying genetic variation related to the functioning of the GH/IGF-1 and mTOR pathways as well as AMPK and Klotho among subsistence populations might similarly reveal trade-offs. For example, do individuals with up-regulated anabolic pathways show greater pathogen defense and reach maturity at younger ages at the expense of accelerated aging? Further, will these individuals be more susceptible to NCD development with lifestyle and environmental shifts?

Research in the developmental origins of health and disease (DOHaD) has highlighted links between early life conditions and later life health (93). For example, both over and under nutrition prenatally predicts increased anabolic health risks (93, 94). Developmental programming of the GH/IGF-1 and mTOR pathways, as well as AMPK and Klotho, might be involved (94, 95). An experimental study of Japanese quail found that increasing prenatal nutrition increased postnatal mTOR and IGF-1 gene expression as well as circulating plasma IGF-1 and body mass (96). Further, in an observational study of human children and adolescents, both higher birthweight and current fat mass percent were positively associated with IGF-1 sensitivity to GH administration (greater increase over 24 hours) (97). These findings are promising for clarifying the mechanisms linking early life conditions to aging and disease. They also raise new questions. For example, while increasing early life calories and nutrients appears to program anabolic pathways toward heightened functioning, what about restriction? Given the link between low birthweight and health risk, does it interact with caloric excess to drive mTOR hyperfunction and disease?

Finally, environmental mismatches are not distributed equally, and behaviors are not devoid of social context, as highlighted by research in the social determinants of health. Access to higher quality foods as well as infrastructure for spaces that allow greater physical activity are unequally dispersed along social class and racial/ethnic lines (98). Individuals of lower social class and marginalized racial/ethnic groups also tend to experience greater psychological stress. The resulting cortisol production can increase cravings for calorically dense foods that activate reward circuitry in the brain and alleviate the stress response (99, 100). There is also evidence that stress-induced depression can reduce motivation for physical activity (101, 102). Further, cigarette smoking and exposure to air pollution is higher among socioeconomically disadvantaged groups (103, 104). These findings highlight the importance of future research on the social determinants of GH/IGF-1, mTOR, AMPK, and Klotho.

**Table 1.** Future research directions

| Why do individuals and species vary along the anabolic-catabolic spectrum, and what are the health consequences? (with corresponding testable hypotheses) | |
|---|---|
| Genetic variation, fitness, and antagonistic pleiotropy | Individuals with upregulated GH/IGF-1/mTOR and downregulated AMPK/Klotho due to genetic variation will be: (1) more resilient to infections in early life, (2) reach maturity earlier, and (3) age faster. |
| Lifestyle shifts and environmental mismatch | Individuals with upregulated GH/IGF-1/mTOR and downregulated AMPK/Klotho due to genetic variation will be more susceptible to NCD development with lifestyle shifts |
| Developmental origins of health and disease | GH/IGF-1, mTOR, AMPK, and Klotho mediate the link between early life metabolic conditions and NCD risk. |
| Social determinants of health disparities | GH/IGF-1, mTOR, AMPK, and Klotho mediate the link between social inequality and NCD risk. |
| Inter-species variation and trade-offs | Species with exceptional longevity for their body size will show relative deficits in anabolic maintenance functions (immune defense and tissue repair/remodeling). |

*Methodological challenges and considerations in humans*

An important caveat and challenge for future population-based studies of mTOR and AMPK is tissue specificity. For example, how informative is mTOR measured in circulation for understanding overall anabolic metabolism? The clustering of anabolic diseases, such as obesity, type 2 diabetes, and autoimmunity, suggests some consistent system-wide differences between individuals. Further, a recent experimental study involving aerobic exercise and intermittent fasting with a sample of obese adults reported a decrease in serum mTOR (105). These observations are suggestive that measuring mTOR and AMPK in circulation can be informative for understanding individual differences in anabolic metabolism and health risk.

Another consideration for measuring mTOR in circulation is whether it fluctuates acutely with an infection due to immune cell proliferation. Future studies are needed to assess this possibility, and it would mean using similar protocols to measures like C-Reactive Protein (CRP) (106). This includes recording participants' infectious symptoms, running sensitivity analyses with these variables, and considering cutoffs for concentrations indicative of an acute infection. Similar to inflammation, understanding health risk will require disentangling acute from chronic low-grade increases in mTOR.

*Why do species vary along the anabolic-catabolic spectrum? What are the trade-offs and health consequences?*

Future research is needed to understand how this anabolic-catabolic functional trade-off operates across species and relates to health and lifespan, similar to what has been proposed as evolutionary gerontology (evo-gero) (107). While most aging research is done with short lived species in laboratory environments, this is increasingly changing. For example, the naked mole rat (NMR) has received considerable attention for being the longest-lived rodent, and there is evidence for the role of anabolic pathway functioning (108). For example, NMR have lower metabolic rates and IGF-1 expression, which could be due to their hypoxic underground environment that creates a nutrient scarce condition (108, 109).

Research on NMR raises the question: what potential costs or trade-offs come with their longevity? There is evidence for a relatively immunosuppressed state. They show a dampened inflammatory response (110), have smaller thymuses than mice, and lack natural killer (NK) cells (111). In addition, NMR cells might be incapable or at least strongly limited in using aerobic glycolysis (the Warburg effect) (112). While this can be protective from tumor growth, it might also attenuate the inflammatory response against pathogens (6). These findings suggest the NMR's longevity might be context specific and only occur with a lower infectious burden, such as in their hypoxic natural environment or a laboratory (111, 113, 114). Future research is needed to understand NMR anabolic maintenance in response to various pathogenic exposures in non-hypoxic environments, as well as cellular repair and remodeling, to understand potential costs.

Bats are also exceptionally long-lived for their body size, and there is genetic evidence suggesting selection on mTOR and autophagy regulation (115). Correspondingly, studies of the bat immune system have shown a dampened inflammatory response and immune system skewed toward tolerance (116-118). This likely explains in part why bats are reservoirs for many viruses, highlighting that lifespan and healthspan are not necessarily coupled. A similar dynamic has been found in the longer lived white-footed deermouse (*Peromyscus leucopus*), which displays a dampened immune response skewed toward tolerance and is a reservoir for a large number of pathogens (119). This tolerance strategy might only work for certain infectious exposures and not necessarily others- for example, the fungus causing white-nose syndrome in bats has shown an extremely high mortality rate. The European greater mouse-eared bat (*Myotis myotis*) is more resilient to white-nose syndrome, which could be due to lower immune tolerance compared to other bat species (120). Similarly to NMR, more research on potential trade-offs and costs that come with longevity in bats can better inform applicability to humans.

**Implications for therapeutics**

*Evolved Compensatory Pathways*

This evolutionary medicine and life history perspective can also provide important insights for therapeutics (**Table 2**). For example, while substantial progress has been made in targeting metabolic pathways for disease prevention, their effects are not as strong as lifestyle factors like reducing caloric excess and physical activity (13, 71). This could be the result of biological degeneracy in metabolic pathways. Degeneracy refers to the ability of different structures to perform similar functions, which is common throughout biological systems (121). Given the critical importance of an organism accurately assessing nutrient availability, a high level of degeneracy in the anabolic-catabolic axis should be expected. In support of this expectation, there are numerous other pathways besides the ones we highlight here, both known and unknown (6, 122).

The implications of degeneracy would mean that drugs targeting one pathway, such as downregulating mTOR, might be less effective in contexts of high caloric consumption and limited physical activity, since other mechanisms might notice the discrepancy in signaling and "correct" the flow of information (**Figure 4**). Degenerate, or alternatively labeled compensatory, pathways have been highlighted in the difficulties of treating multi-drug resistant cancers (123). As a result, there is increasing utilization of multi-target approaches or combining with lifestyle changes (124, 125). The greater efficacy of therapeutics through targeting multiple pathways has been highlighted in neurodegeneration as well, further suggesting degeneracy should be an important consideration for therapeutics (126, 127).

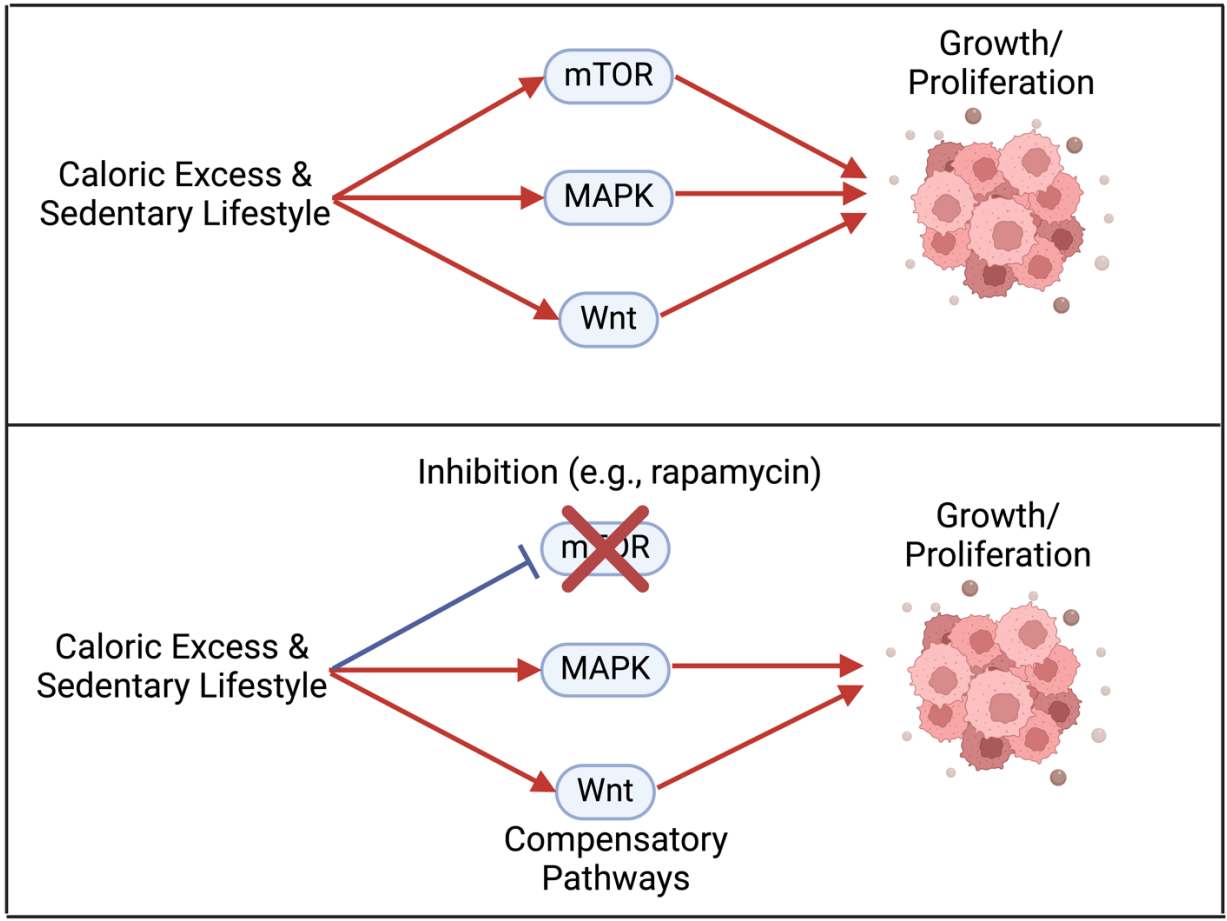

**Figure 4.** Compensatory pathways, such as MAPK and Wnt, that might counteract inhibition of mTOR in the context of caloric excess and sedentary lifestyle, limiting efficacy.

*A Goldilocks zone between anabolic and catabolic functioning*

There is likely a Goldilocks zone for regulating the anabolic-catabolic axis. Too much suppression of growth/proliferation might limit efficacy or even introduce new health risks (**Figure 5**). For example, while moderate caloric restriction has been shown to reverse thymic involution in humans, a marker of immune aging (128), in more extreme restriction there is evidence for compromised function. A study with mice found that while dietary restriction by 40% led to an increased lifespan it decreased the immune repertoire (129). Further, this lifespan extending finding should be interpreted with caution due to the highly controlled laboratory setting. Other studies have found that when mice are exposed to a pathogen in the lab, caloric restriction decreases survival odds (130). Similarly, extreme physical activity in humans, such as endurance training, is commonly found to impair immune function and increase infection risk (131). The health consequences of suppressing anabolic pathways like mTOR too far can also be seen in the original use of rapamycin, which was in high doses as an immunosuppressant for organ transplant recipients. The commonly documented side effect for these patients was increased risk for infections and related cancers (132, 133). Shifting from anabolic to catabolic imbalance could therefore not only increase infection risk but also shift the cancer risk from types that are obesity-related (e.g., breast, colorectal, endometrial, kidney, esophageal, pancreatic, gallbladder) to infection-related (e.g., cervical, Kaposi's sarcoma, lymphoma, certain skin cancers) (134-137). Suggestive evidence for this can be seen among the Tsimane forager-horticulturalists. In addition to their highly active lifestyle and diets of limited caloric excess, they experience frequent helminth infections, which further skews immune function toward tolerance. Correspondingly, while obesity-related cancers are rare, infection-related cancers such as cervical are more common (138).

    A similar shift in etiological profile could also occur in ADRD with catabolic imbalance. While lifestyle factors like caloric excess and limited physical activity contribute to ADRD risk through their effect on cardiometabolic functioning (139, 140), infections and immune function have also been implicated (141, 142). This relationship is complex, as Alzheimer's Disease (AD) might have an autoimmune component (143). As a result, AD risk might shift from one driven by anabolic factors, including lifestyle and autoimmunity, to one driven by catabolic factors, such as

opportunistic infections from impaired immune function. For example, HIV infection, which induces an immunocompromised state, has been associated with elevated dementia risk (144).

The necessity of anabolic maintenance is where Blagosklonny's car analogy becomes problematic, since stopping the car actually appears to increase health risk due to the multifactorial nature of aging (7). Current therapeutics aimed at suppressing mTOR are likely to reveal constraints, in which downregulating the undesirable effect of hyperfunction too far introduces a different undesirable effect: aging due to damage and hypofunction (68). As a result, combinations of therapies will likely be required.

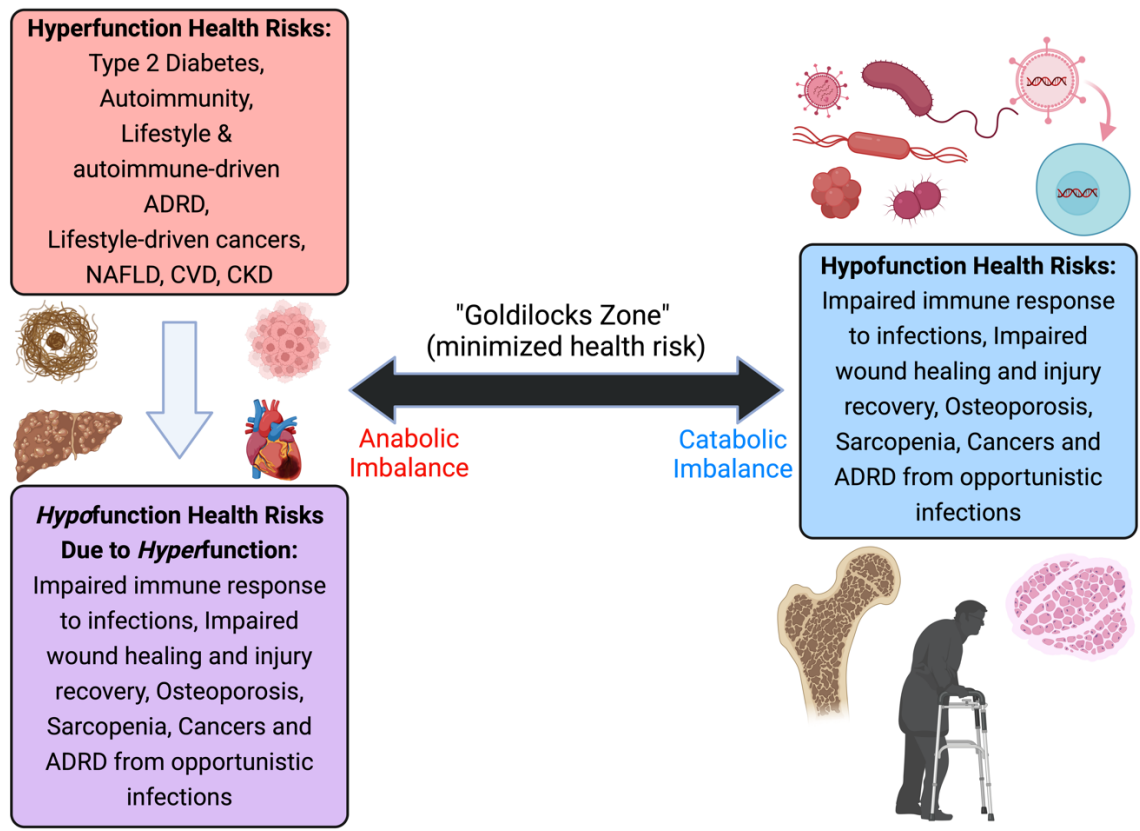

**Figure 5.** The Goldilocks zone of metabolic functioning. While anabolic imbalance is the commonly targeted therapeutic (e.g., mTOR inhibition), too much suppression can lead to catabolic imbalance and its associated hypofunction health risks.

*Species constraints*

A major challenge in aging research has been the translation of findings from short-lived species to longer-lived ones. For example, even under controlled laboratory settings, notable differences have been found between mice and rhesus monkeys. The two major studies on caloric restriction and lifespan in rhesus monkeys provided mixed results, with the Wisconsin study reporting an increased average lifespan and the NIA study reporting no effect (145, 146). Differences in diet and feeding between the studies have been highlighted, as the Wisconsin macaques were fed *ad libitum*, mimicking an overweight human living in an industrialized population (147, 148). In contrast, the NIA study compared a control group that was not fed *ad libitum* with a restricted group. When the studies were combined, they concluded that moderate restriction improves lifespan (149), suggesting limited or no effect for extreme restriction. Studies on caloric restriction are not alone in finding discrepancies between mice and longer-lived primates. For example, a study of Klotho administration found that only low-doses improved cognitive functioning in macaques, while high doses continued to benefit mice (150).

These studies could be indicating species constraints along the anabolic-catabolic axis. Given current evidence for selection on anabolic pathways (57, 58, 108, 115), longer-lived species could already be close to the lower limits of functioning. As a result, extreme anabolic inhibition has limited benefit and can introduce catabolic health risks. The implication of this possibility is that therapeutic interventions aimed at shifting from anabolic to catabolic functioning will continue to show stronger effects in short-lived species like mice but limited effects in humans.

**Table 2.** Implications for therapeutics

| **Compensatory Pathways** | Lifestyle factors of caloric excess and limited physical activity might counteract therapeutic targets (e.g., mTOR inhibition) through compensatory pathways, limiting efficacy. |
|---|---|
| **Goldilocks zone** | The relationship between anabolic metabolism and health risk might be "U" shaped, with too much suppression introducing catabolic health risks such as impaired immune defense. |
| **Species constraints** | Longer lived species might already be near the lower limit of anabolic functioning. As a result, they will benefit less from therapeutics and interventions like caloric restriction, rapamycin, and Klotho compared to shorter lived species. |

**Conclusion**

Here we highlight the value of an evolutionary medicine and LHT perspective for understanding aging and NCD risk. We also highlight connections between the hyperfunction theory of aging and LHT. Central to this perspective is energy utilization, including the functional trade-off between anabolic and catabolic metabolism and energy trade-offs linking hyperfunction and hypofunction. Future research can help clarify how these trade-offs operate across human individuals and across species to understand differences in health. Finally, the multifactorial drivers of aging, including synergies between hyperfunction, hypofunction, and damage, highlight the complexities and challenges for current therapeutic efforts to slow and prevent aging and NCD.

**Acknowledgements**

We thank Mike Gurven and Thom McDade for feedback in the development of this manuscript. Funding support came from the NIH/National Institute on Aging (R01AG054442). Figures were created in BioRender.